\documentclass{article} %Remove draft option to hide ToDos
\include{header}

\title{An Iris for Expected Cost Analysis \\ \normalsize Technical Report}
\author{Janine Lohse, Deepak Garg \\ \normalsize MPI-SWS}
\date{June 02, 2024}

\begin{document}

\renewcommand\cite{\citep}

\maketitle

\hspace{0.1\textwidth} \begin{minipage}{0.8\textwidth}
\small We present \logicName{}, a separation logic framework for the (amortized) expected cost analysis of probabilistic programs. \logicName{} is based on Iris, parametric in the language and the cost model, and supports both imperative and functional languages, concurrency, higher-order functions and higher-order state. \logicName{} also offers strong support for correctness reasoning, which greatly eases the analysis of programs whose expected cost depends on their high-level behaviour. To enable expected cost reasoning in Iris, we build on the expected potential method. The method provides a kind of ``currency'' that can be used for paying for later operations, and can be distributed over the probabilistic cases whenever there is a probabilistic choice, preserving the expected value due to the linearity of expectations. We demonstate \logicName{} by verifying the expected runtime of a quicksort implementation and the amortized expected runtime of a probabilistic binary counter.
\end{minipage}
\vspace{0.3cm}

\pagenumbering{arabic}

\section{Introduction}

The performance of many algorithms relies on randomization. A classical example is Quicksort, whose expected runtime can be improved from $\mathcal{O}(n^2)$ to $\mathcal{O}(n~ log~ n)$ through randomization. However, reasoning about probabilistic algorithms and programs can be counterintuitive, making pen-and-paper proofs error-prone. This highlights the necessity for expressive and mechanized logics for expected cost analysis.

\textit{Iris} \cite{Iris} is an expressive separation logic framework that has been successfully applied to a large number of languages and domains, \eg, \cite{IrisExample1,IrisExample2,IrisExample3,IrisExample4,IrisExample5, IrisExample6}. In this report, we show how to adapt Iris to obtain \textbf{\logicName{}}, a modular and expressive framework for (amortized) expected cost analysis of probabilistic programs. \logicName{} is parametric in both the language and the cost model and applies to feature-rich languages supporting, amongst others, both imperative and functional programming, concurrency, higher-order functions and higher-order state.

We build on the \textit{method of potentials} for amortized cost analysis. Potential can be seen as a ghost ``currency'' that pays for operations that induce cost. Whenever some of the available potential is not immediately needed, it can be stored to pay for future operations. Typically, potential is expressed as function of the program state.

The potential method has been adapted for amortized expected cost of probabilistic programs \cite{AmortizedExp}. The key idea is that when reasoning about expected cost, a probabilistic branching gives the opportunity to \textit{distribute} the potential among the different cases, as long as the expected value of the distributed potentials does not increase. For example, if a program flips a fair coin and a potential of 2 is available, one can continue in the heads-case with a potential of 1 and in the tails-case with a potential of 3. This is sound: if there is a cost of 1 with probability $\frac{1}{2}$, and a cost of 3 with probability $\frac{1}{2}$, the expected cost is 2. 

Existing work has embedded the expected potential method into program logics~\cite{AmortizedExp, ExpAuto, ExpAuto2}. Specifically, \cite{ExpAuto} and \cite{ExpAuto2} are syntax-based approaches designed for automated expected cost analysis. While this approach works great for simple programs, it is not suitable for programs whose expected cost depends on the correctness of their functional behaviour. A crucial reason is the lack of support for correctness reasoning. For example, the expected runtime of any randomized quicksort implementation crucially depends on the \textit{correctness} of the subprocedure implementing the partitioning.
In contrast, \cite{AmortizedExp} supports correctness reasoning, but focuses on a simple imperative language with binary probabilistic choice. It would be very hard in practice to implement and analyze an algorithm like quicksort in the logic of~\cite{AmortizedExp}, since quicksort relies on recursion and uniform sampling from a list.

In contrast, \logicName{} supports all of the following:
\begin{itemize}
    \item We support almost all correctness reasoning that is supported by Iris, with the exception of prophecy variables and later credits.
    \item \logicName{} is parametric in the cost model and language, and supports feature-rich languages, including imperative and functional progamming, concurrency, higher-order functions and higher-order state.
    \item \logicName{} is mechanized in the Coq proof assistant. We also adapt the \textit{Iris Proof Mode} \cite{IPM} for \logicName{}: the Iris Proof Mode provides an interface to keep track of assumptions and goals, and automates routine proof steps. We extend existing tactics for allowing manipulation of potential.
\end{itemize}

We demonstrate our framework with two case studies: We (1) bound the expected runtime of a randomized in-place quicksort implementation and (2) analyze the amortized expected runtime of a probabilistic binary counter.

Our Coq development can be found at \url{https://gitlab.mpi-sws.org/FCS/expiris}.

\subsection{Example} \label{sec:first-example}

To illustrate our method, consider the following program, which simulates tossing a coin until we see a heads, counting a cost of 1 unit for every coin toss:
% \begin{alignat*} {1}
%     &\texttt{toss \_} ~\eqdef~ \Tick 1;;~ 
%                        \If \ChooseUniform \texttt{[heads, tails]} = \texttt{tails} then \texttt{toss} \TT \\
%     &\texttt{toss} \TT 
% \end{alignat*}

\begin{lstlisting}[language=mylang]
    toss _ := tick 1;;
              if ChooseUniform [h, t] = t then toss ()
    toss ()
\end{lstlisting}

In this language, \mcode{tick} is a primitive that increases the incurred cost by 1.  We would like to show that the expected cost of this program does not exceed 2. Specifically, we can show in \logicName{} that after any number of execution steps, the expected cost does not exceed 2.

The coin toss example is a typical example in expected cost analysis. The probability that we have to toss at least once is $\frac{1}{2}$, the probability that we have to toss at least twice is $\frac{1}{4}=\frac{1}{2^2}$, and so on. In total we get 
\begin{align*}
    \mathbb{E}[\mathsf{Tosses}] = \sum_{i=1}^\infty \frac{1}{2^i} = \frac{1}{1 - \frac{1}{2}} = 2 &&\text{(geometric series)}
\end{align*}
In \logicName, we can show the Hoare Triple:
\begin{align*}
    \{\$2, \TRUE\}~ \mcode{toss ()}~\{v.~ 0, v = ()\}   
\end{align*}
which intuitively means the following: given that the precondition $\TRUE$ holds, a potential of 2 is enough to cover the expected cost of executing $\mcode{toss ()}$. Additionally, if the program terminates, there is a potential of at least $0$ left, and the return value is $\mcode{()}$. The postcondition of ``there is a potential of at least $p$ left'' is discussed in more detail in Section \ref{sec:weakestpre}.

The proof proceeds by \ruleref{Löb} induction, which, intuitively, is induction on the number of steps that the program is executed for. It allows us to assume that the statement we want to show already holds for the recursive call, but we have to do at least one program step before applying the induction hypothesis. In \logicName{}, this rule is sound for expected cost analysis even for possibly non-terminating programs like the example discussed in the section (because the postcondition is guaranteed to hold \emph{if} the program terminates). 

The first statement to be executed is a \texttt{tick}, which reduces our (available) potential from 2 to 1. The next statement is a uniform choice. This probabilistic branching allows us to distribute the remaining potential of 1 among the cases. We can choose to give the heads-case a potential of 0 and the tails-case a potential of 2. This is a valid choice, since the expected potential is still $\frac{1}{2} \times 0 + \frac{1}{2} \times 2 = 1$. The proof continues for the two cases separately. After the \texttt{if} is evaluated, we are left with two goals: 
\begin{align*} 
    \{\$0, \TRUE\}~ \mcode{()} ~\{v.~0, v = ()\} \qquad \{\$2, \TRUE\}~ \mcode{toss ()} ~ \{v.~0, v = ()\}
\end{align*}
The first goal is trivially true, and the second one holds by the induction hypothesis. This concludes the proof. 

After proving this triple, we apply \logicName{}'s \emph{adequacy theorem}, which yields that if the program is executed for up to $n$ steps (for any $n$), the expected cost will not exceed 2.

This example is verified formally in Section \ref{sec:formal-example}.

\section{Preliminaries: The Iris Base Logic} \label{sec:base-logic}

Iris is separated into a \textit{base logic} that defines basic connectives and modalities, and a \textit{program logic} that is defined on top of the base logic. We leave the Iris' base logic untouched and only modify the program logic. In this section, we briefly summarize the part of the Iris base logic that is used in this report. 

Iris propositions include the following:
\begin{align*} 
    P, Q \in \mathit{iProp} \bnfdef{}&
    \TRUE \mid \FALSE \mid P \land Q \mid P \lor Q \mid \neg P \mid  
    P \Rightarrow Q \mid \exists x.~ P \mid \forall x.~P \mid \\
    & P \wand Q \mid P * Q \mid \later P \mid \knowInv{\mathcal{N}}{P} \mid \pvs[\mask_1][\mask_2] P \mid  \loc \mapsto v 
\end{align*}

Iris is a \textit{separation logic} \cite{SeparationLogic1, SeparationLogic2}. The base ingredients for a separation logic are the \textit{separating conjunction} $P * Q$ and the \textit{magic wand} $P \wand Q$. The separating conjunction $P * Q$ asserts that the available resources can be separated (linearly) into two parts, one of which satisfies $P$, and the other $Q$. For example, $\ell \mapsto v$ is a resource asserting that the heap location $\loc$ stores the value $v$. A statement $\loc \mapsto v * l' \mapsto v'$ means that the heap can be \textit{separated} into two parts, one satisfying $\loc \mapsto v$, the other satisfying $\loc' \mapsto v'$; in particular, we must have that $\loc \neq \loc'$. 

The wand $P \wand Q$ asserts that if we disjointly add resources that satisfy $P$, then the total resources satisfy $Q$: it holds that $P * (P \wand Q) |- Q$.

The notation $\knowInv{\mathcal{N}}{P}$ describes a sharable \textit{invariant} with namespace $\mathcal{N}$. Typically, the program logic enforces all invariants to hold after every program step: It is possible to \textit{open} an invariant to obtain $P$ but later in the proof, $P$ has to be restored to again, \textit{closing} the invariant.

The fancy update modality $\pvs[\mask_1][\mask_2] P$, allows updating the available resources before proving $P$. The update is needed, for example, for keeping the resources $\loc \mapsto v$ and the physical heap in sync. The fancy update specifically says that $P$ holds in resources obtained \emph{after} (1) opening all invariants contained in $\mask_1$, then (2) updating our resources, and then (3) closing all invariants contained in $\mask_2$. Note that $P$ holds in a state where the invariants in $\mask_1 \backslash \mask_2$ are open and those in $\mask_2$ are closed.

The later modality $\later P$ asserts that $P$ holds after the next program step. The later modality supports \ruleref{Löb} induction, which lifts induction on program steps to the logic:
\begin{mathpar}
    \inferH{Löb}
    {\later P \Rightarrow P}
    {P}
\end{mathpar}

\section{Probabilistic Language} \label{sec:prob-language}

\subsection{Definition}
Next, we adapt Iris' definition of a language to the probabilistic setting.

Iris is language-agnostic: the framework can be instantiated with a language of the user's choice. In Iris, a language consists of \textit{expressions} $\mathit{Expr}$, that describe programs in this language, the \textit{values} $\mathit{Val}$ that a program can evaluate to, and \textit{states} $\mathit{State}$, that can, for example, be used to model the memory state of a machine. The operational semantics of the language is given by a primitive \textit{reduction relation} $ (-, - \step -, -, -) \subseteq (\mathit{Expr} \times \mathit{State}) \times (\mathit{Expr} \times \mathit{State} \times \mathit{List(Expr)})$ that for each pair of expression and state, specifies what steps the program can take next. More specifically, $\expr, \state, \step \expr', \state', \vec{\expr}$ means that the expression $\expr$ can reduce to $\expr'$, changing the state to $\state'$ and forking the expressions in $\vec{\expr}$ as new threads. For example, in a simple language, for an arbitrary state $\state$ and expressions $\expr, \expr'$, the following three reduction triples might hold:
\[
\begin{array}{rcl}
    \mcode{if true then }~ e ~\mcode{ else }~ e', \state  & \step & e, \state, [] \\
    l \leftarrow 5, [l \mapsto 3]  & \step & (), [l \mapsto 5], [] \\
    \mcode{fork }~ \expr, \state & \step & (), \state, [\expr]
\end{array}
\]
%This kind of program semantics is \textit{small-step}, because it models individual computation steps instead of directly specifying which value an expression evaluates to. 

For cost analysis, we modify this description of languages. First, the description also specifies the cost of each step. Second,
it is necessary to capture the probability distribution of possible next steps. Therefore, we modify the stepping relation so that a pair of expression and state now steps to a distribution over tuples of expression, state, cost and forked expressions. This leads to a new form of primitive reduction relation.

\begin{definition}[Discrete distribution] 
    For a countable set $X$, a \textit{distribution} over $X$ is a function $\distr \in (X \rightarrow \mathbb{R}_{\geq 0})$ that satisfies $\sum_{x \in X} \distr(X) = 1$. 
     The support of a distribution is the set of outcomes with nonzero probability: $\supp{\distr} = \{x \in X \mid \distr(x) > 0\}$.
\end{definition}
We are only interested in discrete distributions with \emph{finite support} and we write $\mathbb{D}(X)$ for the set of such distribution over $X$.

\begin{definition}[Probabilistic Language]
    A \textit{probabilistic language} is a tuple \\ $(\mathit{Expr}, \mathit{Val}, \mathit{State}, \step)$, where
    $\mathit{Expr}$ is the set of expressions, $\mathit{Val} \subseteq \mathit{Expr}$ is the set of values, $\mathit{State}$ is the set of possible states, and 
    \begin{align*}
        (-, - \step -) \subseteq (\mathit{Expr} \times \mathit{State}) \times \mathbb{D}(\mathit{Expr} \times \mathit{State} \times \mathbb{R}_{\geq 0} \times \mathit{List(Expr)})
    \end{align*}
    is the primitive reduction relation. Additionally, values must not step:
    \begin{align*} 
        \forall \expr, \state, \distr.~ \expr, \state \step \distr \Rightarrow \expr \notin \mathit{Val}
    \end{align*}
\end{definition}

For example, an expression $flip$ that simulates a fair coin flip, could step to the value $heads$ with probability $\frac{1}{2}$ and to the value $tails$ with probability $\frac{1}{2}$, and incur a cost of 1:
\begin{align*}
    \mcode{flip}, \state \step \begin{cases}
        \mcode{heads}, \state, 1, [] &\text{with probability } \frac{1}{2} \\
        \mcode{tails}, \state, 1, [] &\text{with probability } \frac{1}{2}
    \end{cases}
\end{align*}

Next, we show how to combine the individual steps into a relation modeling $n$ steps of program execution. This will be needed for stating the adequacy theorem (see Section~\ref{sec:adequacy}). Adequacy states, amongst others, that proving a weakest precondition in \logicName{} ensures that the expected cost does indeed not exceed the stated bound after $n$ steps of program execution (for any $n$).

Since we support concurrency, the reduction relation needs to be generalized from individual expressions to thread pools. We follow the Iris approach of interleaving thread reductions arbitrarily so, at any point in time, any thread may step next. Additionally, we now allow starting with a nonzero cost:
\begin{align*}
    (-, -, - \rightarrow_{tp} -) \subseteq (List(\Expr) \times State \times \mathbb{R}_{\geq 0}) \times \mathbb{D}(List(Expr) \times State \times \mathbb{R}_{\geq 0})
\end{align*}
\begin{align*}
\infer{e, \sigma \rightarrow_t \distr}{\tpool \dplus [\expr] \dplus \tpool', \sigma, \cost \rightarrow_{tp} map(\distr, \lambda (\expr', \sigma', \cost', \vec{e}).~(\tpool \dplus [\expr'] \dplus \tpool' \dplus \vec \expr), \sigma', \cost + \cost')} \\
\text{where } map(\distr, f)(y) := \sum_{x \in f^{-1}(y)}  \distr(x)
\end{align*}

Next, we define a relation, denoted $\rightarrow_{tp}^n$, modeling up to $n$ steps of program execution. For composing multiple steps, we apply (a special case of) the \textit{bind} operation of the Giry Monad \cite{Giry} of distributions.
\begin{align*}
bind : \mathbb{D}(X) \rightarrow (X \rightarrow \mathbb{D}(Y)) \rightarrow \mathbb{D}(Y) \\
\bind{\distr}{\distrf}(y) = \sum_{x \in X}  \distr(x) \cdot \distrf(x)(y)
\end{align*}
Intuitively, for obtaining $y$ after at most $n$ steps, we first have to step to some $x$ in step 1, and then from $x$ to $y$ with at most $n-1$ additional steps. To obtain the total probability of reaching $y$, we sum up the probabilities for all possible intermediate steps $x$.
\begin{align*}
    (-, -, - \rightarrow_{tp}^n -) \subseteq (List(Expr) \times State \times \mathbb{R}_{\geq 0}) \times \mathbb{D}(List(Expr) \times State \times \mathbb{R}_{\geq 0})
\end{align*}
\begin{align*}
\infer{}{\tpool, \state, \cost \rightarrow_{tp}^n \dirac{(\tpool, \state, \cost)}} & ~~~~~(n \geq 0)
\end{align*}
\begin{align*}
\infer{\tpool, \state, \cost \tpstep \distr \qquad \forall (\tpool', \state', \cost') \in \supp{\distr}.~ T', \state', \cost' \tpstep^{n-1} \distrf((T', \state', \cost'))}{\tpool, \state, \cost \tpstep^n \bind{\distr}{\distrf}}
\end{align*}

\subsection{Example Language: Probabilistic Heap Lang}

We instantiate the framework above with a concrete probabilistic language by extending Iris' example language HeapLang for probabilistic programming. HeapLang is a simple but expressive language supporting concurrency, both imperative and functional programming, higher-order state and programs. We make the following extensions to the language:
\begin{itemize}
    \item We add support for the manipulation of reals (to model costs and probabilities).
    \item We add the new primitive $\mcode{tick}$ that increases the cost by the specified amount. 
    \item We add the probabilistic primitives $\mcode{ChooseUniform}$ and $\mcode{ChooseWeighted}$.\\ $\mcode{ChooseUniform}$ uniformly samples a value from a given list. $\mcode{ChooseWeighted}$ samples a value from a list of pairs, where each pair contains a value and a real number. The number represents the weight of this value. The weights are normalized to sum to 1 and interpreted as probabilities, and a value is sampled.
    \item We add a shallow embedding of Coq lists to HeapLang to simplify working with the new probabilistic primitives.
\end{itemize}

The following grammar defines the syntax of ProbHeapLang, where $z \in \integer$, $r \in \mathbb{R}$, $\loc \in \Loc$ and $vs \in List~ \Val$:
\begin{align*}
\val,\valB \in \Val \bnfdef{}&
  z \mid r \mid vs \mid
  \mcode{true} \mid \mcode{false} \mid
  \mcode{()} \mid
  \loc \mid
  {}&  \\&
  \mcode{rec}_v~ \lvarF(\lvar)= \expr \mid
  (\val,\valB)_\valForm \mid
  \InlV(\val) \mid
  \InrV(\val)  \\
\expr \in \Expr \bnfdef{}&
  \val \mid
  \lvar \mid
  \RecE\lvarF(\lvar)= \expr \mid
  \expr_1(\expr_2) \mid
  {}\\ &
  - \expr \mid 
  \expr_1 + \expr_2 \mid \expr_1 - \expr_2 \mid \expr_1 \cdot \expr_2 \mid \expr_1 / \expr_2 {} \mid e_1 :: e_2 \\ &
  \expr_1 < \expr_2 \mid \expr_1 \leq \expr_2 \mid \expr_1 = \expr_2 \mid 
  \expr_1 \land \expr_2 \mid \expr_1 \lor \expr_2 \mid
  \If \expr then \expr_1 \Else \expr_2 \mid
  {}\\ &
  (\expr_1,\expr_2)_\exprForm \mid
  \Fst(\expr) \mid
  \Snd(\expr) \mid
  {}\\ &
  \InlE(\expr) \mid
  \InrE(\expr) \mid
  \Match \expr with \Inl => \expr_1 | \Inr => \expr_2 end \mid
  {}\\ &
  \AllocN(\expr_1,\expr_2) \mid
  \Free(\expr) \mid
  \deref \expr \mid
  \expr_1 \gets \expr_2 \mid {}\\ &
  \CmpXchg(\expr_1, \expr_2, \expr_3) \mid
  \Xchg(\expr_1, \expr_2) \mid
  \FAA(\expr_1, \expr_2) \mid
  \kern-30ex{}\\ &
  \Fork \expr \mid \\&
  \Tick \expr \mid \ChooseUniform \expr \mid \ChooseWeighted \expr
\end{align*}

The semantics is mostly standard. Non-probabilistic primitives step to a Dirac distribution of the form $\delta_x$, which assigns the value $x$ the probability 1. $\deref \loc$ reads location $\loc$ and $\loc \gets v$ stores $v$ in location $\loc$. \texttt{CmpXchg} is an atomic compare-and-store primitive, while \texttt{Xchg} is an atomic load-and-store. \texttt{FAA}$(\loc, z)$ is an atomic primitive that reads $\loc$ and adds $z$ to its value. 

The reduction rules for the new constructs -- \texttt{tick} and the probabilistic primitives -- are as follows.
\begin{align*} 
    \infer{r \in \mathbb{R}_{\geq 0}}
    {\Tick r, \state \step \dirac{(\TT, \state, r, [])}} \qquad 
    \infer{vs \neq [] \qquad \forall v \in vs.~ \distr((v, \state, 0, [])) = \frac{1}{length(vs)}}
    {\ChooseUniform vs, \state \step \distr}
\end{align*}
\begin{align*}
    \infer{vs \neq [] \qquad \forall v \in vs.~ v = (r, v')_\valForm \wedge r \in \mathbb{R}_{+} \wedge \distr((v', \state, 0, [])) = \frac{r}{s} \qquad s = \sum_{(r, v') \in vs} r}{\ChooseWeighted vs, \state \step \distr }
\end{align*}

\newpage

\section{Probabilistic Weakest Precondition} \label{sec:weakestpre}
	In Iris, the heart of the program logic is the \textit{weakest precondition}. Proving a weakest precondition $\wpre e {\pred}$ shows that $e$ executes safely and if it terminates in a value $v$, $v$ fulfills the \textit{postcondition} $\pred$. 
	
	To enable expected cost analysis, we extend the weakest precondition with two new parameters: the initial potential $\pota \in \mathbb{R}$ and the postcondition potential $\potb \in Val \rightarrow \mathbb{R}$. Then, the weakest precondition 
	\begin{align*}
		\wpP{\pota}{\expr}{\potb, \pred} \qquad \text{or equivalently} \qquad \wpP{\pota}{\expr}{v.~\potb(v), \pred(v)}
	\end{align*}
	 intuitively means that the initial potential $\pota$ is enough to cover the expected cost of executing $\expr$ and additionally, if $e$ terminates in a value $v$, we still have a potential of $\potb(v)$ left. Additionally, as in standard Iris, safe execution of $e$ and the fulfillment of the postcondition $\pred$ are guaranteed.
	
	In particular, proving $\wpP{\pota}{\expr}{\potb, \pred}$ ensures that the expected cost of executing $\expr$ does not exceed $\pota$. The postcondition potential $\potb$ enables modular proofs. To see why it is useful, consider the following example: 
	%
	% \begin{alignat*} {1}
	% 	&\texttt{toss \_} ~\eqdef~
	% 					   \If \ChooseUniform \texttt{[heads, tails]} = \texttt{tails} then \texttt{toss} \TT \\
	% 	&\Tick (\texttt{toss} \TT)
	% \end{alignat*}
	%
	\begin{lstlisting}[language=mylang]
	toss _ := if ChooseUniform [true, false] then 1
			else 1 + toss ()
	Tick(toss ())
	\end{lstlisting}
	This is a variant of the coin toss example discussed in Section \ref{sec:first-example}. Here, we first count the number of coin tosses and then incur a cost of this amount. To prove that this program has an expected cost of at most 2, we can first prove the weakest precondition $\wpP{2}{\mcode{toss()}}{v.~ v, \TRUE}$ by induction. This shows that if we execute $\mcode{toss()}$ with an initial potential of 2, we can distribute this potential in a way such that if we terminate with a value $v$, then the final potential is also $v$. Next, we also show that $\wpP{v}{\Tick v}{v.~0, \TRUE}$, which directly follows from the rule \ruleref{wp-tick}. Then, the two parts of the proof are connected using the rule \ruleref{wp-bind}:  
	\begin{mathpar} 
	\inferH{wp-bind}
	{\wpP \pota \expr {v.~\potb'(v), \wpP {\potb'(v)} {K[v]} {\potb, \pred}}}
	{\wpP \pota {K[e]} {\potb, \pred}}  \qquad
	\inferH{wp-tick}
		{\later \pred~ ()}
		{\wpP c {\Tick c} {\mathbf{0}, \pred}}
	\end{mathpar}
	The rule \ruleref{wp-bind} allows us to reason about a composed expression by reasoning about both parts sequentially. The rule says that we should first evaluate $\expr$ with the initial potential $\pota$, yielding value $v$ and potential $\potb'(v)$, and then continue the proof for $K[v]$ with the potential that remains in for this case. For proving the example above, we can instantiate the rule in the following way:
	\begin{mathpar}
	\inferH{ }{\wpP {2} {\texttt{toss ()}} {v.~v,\wpP {v} {\Tick(v)} {0, \TRUE}}}{\wpP {2} {\Tick \texttt{toss ()}} {0, \TRUE}}
	\end{mathpar}
        The premise of the rule follows from the two triples we discussed earlier.
        
	The weakest precondition further satisfies the following rules:
	\begin{mathpar}
		\inferH{wp-value}
		{\potb(v) \leq \pota \qquad \pred~ v}{\wpP \pota v {\potb, \pred}}
		\end{mathpar}
	\begin{mathpar}
		\inferH{wp-add-pot}
		{\pota' \geq 0 \qquad \wpP \pota \expr {\potb, \pred}}
		{\wpP {(\pota + \pota')} \expr {\lambda v.~ \potb v + \pota', \pred}} \qquad
		\inferH{upd-wp}
		{\pvs \wpP p \expr {\potb, \pred} }
		{\wpP p \expr {\potb, \pred}}
		\end{mathpar}
	\begin{align*}
		\inferH{wp-mono}
		{\pota \leq \pota' \qquad \forall v.~ \potb v \geq \potb' v \qquad \forall v.~ \pred v \wand \Psi v \qquad \wpP \pota \expr {\potb, \pred}}
		{\wpP {\pota'} \expr {\potb', \Psi}}
	\end{align*}

	A \textit{pure} step is a deterministic computation step that has no side effects: $e \rightarrow_{pure} e' \eqdef \forall \state.~ e, \state \step \dirac {(e', \state, 0, [])} $. The weakest precondition is closed under the inverse of pure reductions. 
	\begin{mathpar} 
		\inferH{wp-pure}
		{\wpP{\pota}{\expr'}{\potb, \pred} \qquad \expr \rightarrow_{pure} \expr'}
		{\wpP{\pota}{\expr}{\potb, \pred}}
	\end{mathpar}

	Additionally, the following rules hold for primitives of ProbHeapLang:
        
	\begin{mathpar}
		\inferH{wp-choose-uniform}
		{l \neq [] \qquad \forall v \in l.~ \pred~ v \qquad \sum_{v \in l} \potb(v) \leq \pota \cdot length~ l }
		{\wpP {\pota} {\ChooseUniform{l}} {\potb, \pred}} \qquad 
	\end{mathpar}
	If there is a uniform choice in the program, the postcondition potential $\potb$ specifies how to distribute the potential $p$ among the fibers. The condition $\sum_{v \in l} \potb(v) \leq \pota \cdot length~ l$ ensures that the expected available potential does not increase. Moreover, we have to show the postcondition for each value in the list.

	\begin{mathpar}
		\inferH{wp-choose-weighted}
		{vs \neq [] \qquad \forall v \in vs,~ \exists r, v'.~ v = (r, v') * r > 0 * \Phi~ v' \qquad \frac{\sum_{(r, v) \in vs} r \cdot \potb (v)}{\sum_{(r,v) \in vs} r} \leq p}
		{\wpP \pota {\ChooseWeighted vs} {\potb, \Phi}} 
	\end{mathpar}
	For a weighted choice, we have to prove that the list in input indeed consists of pairs of reals and values that fulfill the postcondition. The premise $$\left( \sum_{(r, v) \in vs} r \cdot \potb (v)\right) / \left(\sum_{(r,v) \in vs} r \right) \leq p$$ ensures that the expected value of the available potential does not increase. (The denominator normalizes the sum of the weights 1.)
        
	\begin{mathpar}
		\inferH{wp-fork}
		{\later \pred~ () \qquad \later \wpP p e {\mathbf{0}, \TRUE}}
		{\wpP p {\Fork e} {\mathbf{0}, \pred}}
	\end{mathpar}
	When a new thread is forked, it can use up the full available potential.

	In addition to the above rules, all rules for the primitives from HeapLang (except for prophecy) hold in \logicName{} if we define  the standard Iris precondition as:
        \[\wpre {\expr} {\pred} \eqdef \wpP {0} {\expr} {\mathbf{0}, \pred}\]
        Those rules can be generalized to any arbitrary nonnegative potential using the rule \ruleref{wp-add-pot}. Moreover, proofs of HeapLang programs in Iris + HeapLang are also \logicName{} + ProbHeapLang proofs as long as they don't make use of prophecies and later credits. This allows us to use the existing library of verified HeapLang programs when carrying out proofs in ProbHeapLang.

	\subsection{Recap of the Iris Weakest Precondition Definition}

	For the purposes of our exposition, we show a slightly simplified definition of the standard Iris weakest precondition $\wpre {\expr} {\pred}$. We will adapt this definition to \logicName{} in the next section. The full definition is a bit more expressive: it additionally allows the state interpretation to depend on the step counter and the number of threads, it is parametric in whether the program is allowed to get stuck, parametric in the postcondition that should hold for forked threads, and may contain more laters in every step.

	Additionally, this definition does not include prophecy variables or later credits. A line-by-line explanation can be found below.
	\begin{align}
		\textdom{wp}(\stateinterp) \eqdef{}& \MU \textdom{wp\any rec}. \Lam \mask, \expr, \pred. \\
        & (\expr \in Val \land \pvs[\mask] \pred(\expr) ) \lor {}\\ 
		& \Bigl(\expr \notin Val \land \All \state. \stateinterp(\state)  \wand \pvs[\mask][\emptyset] {}  \red(\expr, \state) *  \\
        & \All \expr', \state', \vec{\expr}.~ (\expr , \state  \step \expr', \state', \vec{\expr} ) \wand \\
        &\later \pvs[\emptyset][\mask] \stateinterp(\state') * \textdom{wp\any rec}(\mask, \expr', \pred) * {}\\
		&\qquad \Sep_{\expr'' \in \vec\expr} \textdom{wp\any rec}(\top, \expr'', \TRUE) \Bigr)
	\end{align} 

	\begin{enumerate}
		\item The weakest precondition is parametric in the \textit{state interpretation} $\stateinterp$. The state interpretation connects the physical state to the logic. It has to be preserved by every program step. For more details on how the state interpretation is used, we refer to Iris \cite{Iris}.
		\item If the program has terminated in a value, the postcondition must hold for that value.
		\item Next, consider the case that the program has not yet reduced to a value. The machine could now be in any state $\state$ that fulfills the state interpretation. First, one must show that the expression is reducible, which means that the program should not be stuck. For this part of the proof, the invariants can stay open.
		\item Next, consider all possible combinations of expression, state, and forked expressions that the program could step to.
                \item Since the program reduced one step, the remaining formula is under a later modality. At this point, one has to close the invariants again, and can also update the resources. The state interpretation must hold on the new state, and the weakest precondition must hold recursively for the next expression.
		\item All forked threads should execute safely.
	\end{enumerate}

	\subsection{Definition of the \logicName{} Weakest Precondition} \label{sec:weakestpre-def} 
	\setcounter{equation}{0}

	Next, we turn to the formal definition of the \logicName{} weakest precondition. Compared to the current Iris version, we drop support for prophecy and later credits.
	Additionally, the definition shown in this section is simplified in the same way as the Iris weakest precondition was simplified in the previous section. For the full definition, see Appendix \ref{wp-full}

	When an expression steps to a distribution $\distr$ in \logicName{}, one has to choose how to distribute the available potential among the fibers of $\distr$. To avoid unnecessary side conditions on the functionality of this potential distribution, we allow the user to give the potential distribution as a list $\vec{p}$. For defining the expected value of such a list, we assume an arbitrary order on the support of $\distr$ and define for $n=|\supp{\distr}| = length(\vec{p})$: $\mathbb{E}_\distr[\vec{p}] = \sum_{i=1}^n \distr(\supp{\distr}_i) \cdot \vec{p}_i$.
	After the potential has been distributed, we have to (recursively) show a weakest precondition for each fiber in its allocated potential.

	The \logicName{} weakest precondition is shown below with changes relative to the standard Iris weakest precondition are highlighted in \highlight{yellow}. A line-by-line explanation of the changes can be found below.
	\begin{align}
		\textdom{wp}(\stateinterp) \eqdef{}& \MU \textdom{wp\any rec}. \Lam \mask, \expr, \pred, \highlight{\pota, \potb}. \\
        & (\expr \in Val \land \pvs[\mask] \pred(\expr) \wedge \highlight{\pota \geq \potb(\expr)}) \lor {}\\ 
		& \Bigl(\expr \notin Val \land \All \state. \stateinterp(\state)  \wand \pvs[\mask][\emptyset] {}  \red(\expr, \state) * \highlight{\pota \geq 0} * \\
        & \highlight{\All \distr. (\expr , \state  \step \distr ) \wand} 
        \later\pvs[\emptyset] \highlight{\exists \vec{\pota}.~\mathbb{E}_{\distr}[\vec{\pota}] \leq \pota - \mathbb{E}_{(\_, \_, c) \sim \distr}[c]} * \\
		&\highlight{\bigwedge\limits_{p'; (\expr', \state', \_ , \vec{\expr}) \in \vec{p}; \supp \distr} \Bigl(
			\exists \pota'', \vec{pt}.~ \pota'' + \sum \vec{pt} \leq \pota' ~ \wedge } \\
		&\qquad  \pvs[\emptyset][\mask] \stateinterp(\state') * \textdom{wp\any rec}(\mask, \expr', \pred, \highlight{\pota'', \potb}) * {}\\
		&\qquad \Sep_{\expr''; \pota''' \in \vec\expr; \vec{pt}} \textdom{wp\any rec}(\top, \expr'', \TRUE, \highlight{\pota''', 0}) \Bigr)\Bigr) 
	\end{align} 

	\begin{enumerate}
		\item  The weakest precondition definition takes the initial and final potentials as additional parameters.
		\item When $e$ is a value, in addition to proving the postcondition $\pred$, we have to show that the initial potential is at least as large as the final postcondition.
		\item The potential should never become negative. This is a standard requirement in the method of potentials.
		\item We now have to consider all possible distributions that the expression can step to. Fix such a distribution $\distr$. The expected cost $\mathbb{E}_{(\_, \_, c) \sim \distr}[c]$ incurred by this execution step is deducted from the potential, and we have to choose how to distribute the remaining potential among the fibers of $\distr$. This later distribution is given by a list of potentials $\vec{p}$, which must have one entry for every fiber. This can be done in any way, as long as the expected value does not exceed the remaining potential. The invariants can remain open up to this point.
		\item Once the potential is distributed, we reason separately for each fiber (the big $\bigwedge$), and the distribution $\distr$ becomes irrelevant. Consider a concrete fiber with an assigned potential of $p'$. If new threads were forked, we have to decide how to distribute the potential among all threads. $p''$ is the potential for the main thread, and $\vec{pt}$ are the potentials for the forked threads. Again, these can be chosen arbitrarily as long as their sum is at most $p'$.
		\item The state interpretation has to hold on the new state, and the proof continues recursively for the reduct expression and the initial potential $p''$.
		\item All forked threads should execute safely with the potentials $\vec{pt}$ assigned to them.
	\end{enumerate}
	
	We define the following notation:
	\begin{align*}
		\wpP{\pota}[\stateinterp]{\expr}{\potb, \pred} \eqdef& ~\textdom{wp}(\stateinterp)~ \top ~\expr~\pred~\pota~\potb \\
		\wpre[\stateinterp]{\expr}{\pred} \eqdef&
		 ~\wpP {0}[\stateinterp]{\expr}{\mathbf{0}, \pred} 
	\end{align*}

        When the state interpretation $\stateinterp$ is unambiguously fixed in the context, we often omit it from  $\wpP{\pota}[\stateinterp]{\expr}{\potb, \pred}$ and simply write $\wpP{\pota}{\expr}{\potb, \pred}$. We then define:
        \begin{align*}
		 \{\$p, P\}~ e ~\{\potb, \Psi\} \eqdef&~ P |- \wpP {p} e{\potb, \Psi}
        \end{align*}
        
	\subsection{Example} \label{sec:formal-example}

We now formally verify the example that was already discussed at a high level in Section \ref{sec:first-example}. 
%
%\begin{alignat*} {1}
%    &\texttt{toss \_} ~\eqdef~ \Tick 1;;~ 
%                       \If \ChooseUniform \texttt{[h, t]} = \texttt{t} then \texttt{toss} \TT \\
%    &\texttt{toss} \TT 
%\end{alignat*}
%
	\begin{lstlisting}[language=mylang]
toss _ := tick 1;;
	  if ChooseUniform [h, t] = t then toss ()
toss ()
	  \end{lstlisting}
	
	Recall that we want to prove that $\{\$2, \TRUE\}~ \texttt{toss} () ~\{v.~ 0, v = ()\}$. A formal proof outline is shown below. The initial goal is at the top of the diagram, and the following steps are a (linearized) representation of the proof tree.

	\begin{mathpar}
            \begin{spfoutline}
                \pfhaveshow{}{\{\$2, \TRUE\}~ \texttt{toss()} ~\{v.~ 0, v = ()\}}
				\hline 
				\pfshow{\wpP{2}{\texttt{toss()} }{v.~0, v=()}}
				\hline
				\pfhaveshow{\later \wpP{2}{\texttt{toss ()}}{v.~0, v=()}}{\wpP{2}{\texttt{toss()}}{v.~0, v=()}}
				\pftext{by Löb induction.}
				\hline 
				\pfhave{\wpP{2}{\texttt{toss ()}}{v.~0, v=()}}
				\pfshow{\wpP{2}{\Tick 1;;
				         \If \ChooseUniform \mcode{[h, t] = t} then \texttt{toss()}}{v.~ 0, v = ()}}
				\pftext{by \ruleref{wp-pure}.}
				\hline
				\pfhave{\wpP{2}{\texttt{toss()}}{v.~0, v=()}}
				\pfshow{\wpP{2}{\Tick 1}{v'.~1,
				         \wpP {1} {v';;\If \ChooseUniform \texttt{[h, t] = t} then \texttt{toss()}}{v.~ 0, v = ()}}}
				\pftext{by \ruleref{wp-bind}.}
				\hline 
				\pfhave{\wpP{2}{\texttt{toss()}}{v.~0, v=()}}
				\pfshow{\wpP{1}{\Tick 1}{v'.~0, 
				         \wpP {1} {v';;\If \ChooseUniform \texttt{[h, t] = t} then \texttt{toss()}}{v.~ 0, v = ()}}}
				\pftext{by \ruleref{wp-add-pot}}
				\hline 
				\pfhave{\wpP{2}{\texttt{toss()}}{v.~0, v=()}}
				\pfshow{{\wpP {1} {();;\If \ChooseUniform \texttt{[h, t] = t} then \texttt{toss()}}{v.~ 0, v = ()}}}					
				\pftext{by \ruleref{wp-tick}, \ruleref{wp-value}}
				\hline 
				\pfhave{\wpP{2}{\texttt{toss()}}{v.~0, v=()}}
				\pfshow{{\wpP {1} {\If \ChooseUniform \texttt{[h, t] = t} then \texttt{toss()}}{v.~ 0, v = ()}}}	
				\pftext{by \ruleref{wp-pure}.}
				\hline 
				\pfhave{\wpP{2}{\texttt{toss()}}{v.~0, v=()}}
				\pfshow{\wpP {1} {\ChooseUniform \texttt{[h, t]}}{v'.~ \potb(v'), \wpP {\potb(v')} {\If v' = \texttt{t} then \texttt{toss} ()}{v.~ 0, v = ()}}}	
				\pftext{by \ruleref{wp-bind} with $\potb(h) = 0$, $\potb(t) = 2$.}
				\hline
				\pftext{apply \ruleref{wp-choose-uniform}}
				\hline \hline 
				\pftext{Subgoal 1}
				\pfshow{\sum_{v' \in [h, t]} \potb(v) \leq 1 \cdot length([h, t])}
				\pfshow{0 + 2 \leq 2}
				\pfshow{\checkmark}
				\hline \hline 
				\pftext{Subgoal 2}
				\pfhave{\wpP{2}{\texttt{toss()}}{v.~0, v=()}}
				\pfshow{\forall v' \in [h, t].~ \wpP {\potb(v')}
				{\If v' = \texttt{t} then \texttt{toss()}}{v.~ 0, v = ()}}
				\hline 
				\hline 
				\pftext{Subgoal 2.1}
				\pfshow{\wpP{0}{\If \texttt{h=t} then \texttt{toss()}}{v. 0, v=()}}
				\hline 
				\pfshow{\wpP{0}{()}{v. 0, v=()}}
				\pftext{by \ruleref{wp-pure}}
				\hline 
				\pfshow{0 \leq 0 \wedge () = ()}
				\pfshow{\checkmark}
				\pftext{by \ruleref{wp-value}}
				\hline 
				\hline 
				\pftext{Subgoal 2.2}
				\pfhave{\wpP{2}{\texttt{toss()}}{v.~0, v=()}}
				\pfshow{\wpP{0}{\If \texttt{t=t} then \texttt{toss()}}{v. 0, v=()}}
				\hline 
				\pfhave{\wpP{2}{\texttt{toss()}}{v.~0, v=()}}
				\pfshow{\wpP{2}{\texttt{toss()}}{v.~0, v=()}}
				\pfshow{\checkmark}
				\pftext{by \ruleref{wp-pure}, Assumption.}
				\hline
            \end{spfoutline} 
        \end{mathpar}  

	While this proof is quite verbose on paper, it is only three lines of code within the Coq formalization (instead of values \texttt{heads, tails} which don't exist in ProbHeapLang, we use \mcode{true} and \mcode{false}).

	\begin{lstlisting}
Proof.
  wp_pures. iLob as "IH". wp_tick. iNext. wp_pures.
  wp_choose_uniform (fun v, match v with | #true => 2 | _ => 0 end).
  repeat iSplit; (auto || by wp_pures).
Qed.
	\end{lstlisting}

	The routine application of the rules \ruleref{wp-bind} and \ruleref{wp-add-pot} is handled automatically by our tactics \texttt{wp\_tick} and \texttt{wp\_choose\_uniform}. For the uniform choice, the user only has to provide a function that specifies how to distribute the potential. Since the proof showing that the expected value does not increase is trivial in this case, it is also handled automatically. \ruleref{wp-value} is also applied automatically whenever possible.

\section{Adequacy} \label{sec:adequacy}
	
	\newcommand\metaprop{\varphi}
	\newcommand\consstate{C}
	
	The adequacy theorem says that proving a weakest precondition implies the following three statements:
	\begin{enumerate}
		\item The postcondition $\pred$ holds for all threads that have terminated in values.
		\item Progress: the program does not get stuck. If it terminates, it does so in a value.
		\item The expected sum of the final cost and the postcondition potential does not exceed the sum of the initial potential and the initial cost. In other words, the expected cost of the verified program, which is the difference between the expected final cost and the initial cost does not exceed the difference between the expected final potential and the initial potential.
                  %% This also implies that the expected cost is bounded by the sum of initial potential and initial cost. 
                  The expected sum of cost and postcondition potential for a distribution $\distr$ and a potential postcondition $\potb$ is expressed by the following function $\pcost{\distr}{\potb}$. 
		    \begin{align*} 
			\distr&: \mathbb{D}(List(Expr) \times State \times \mathbb{R}) \\
			Cost_{\potb}((\expr_1 \dplus \vec{\expr}, \state, \cost)) &\eqdef 
			\begin{cases}
				\cost + \potb(\expr_1) & \text{if } \expr_1 \in Val \\
				\cost & \text{else.}
			\end{cases} \\
			\pcost{\distr}{\potb} &\eqdef \mathbb{E}_{\distr}[Cost_{\potb}]
		\end{align*}
		Note that the potential postcondition only applies to the main thread, and thus $\vec{\expr}$ is irrelevant in $Cost_{\potb}((\expr \dplus \vec{\expr}, \state, c))$.
	\end{enumerate}

	All proofs of (1)--(3) above proceed by induction over $\expr, \state, \cost \tpstep^n \distr$.
	For (3), the induction step follows from a variant of the law of total expectation, that allows us to compose the expected cost bounds for each fiber into a bound for the whole distribution:

	\begin{lemma}
		\begin{align*}
		\left(\forall \pota';(\expr, \state, \cost) \in \vec{\pota}; \supp \distr.~ \pcost{\distrf (\expr, \state, \cost)}{\potb} \leq p' + c\right) \rightarrow  \\
		\pcost{bind(\distr, \distrf)}{\potb} \leq \mathbb{E}_{(\_, \_, c) \sim \distr}[c] + \mathbb{E}_\distr[\vec{p}]
		\end{align*}
	\end{lemma}

	The three parts are, together with the soundness result of the Iris' base logic, composed into a full adequacy theorem, which can be found in Appendix \ref{adequacy_full}. For simplicity, we only present a slightly weaker corollary in this section: Proving a weakest precondition and the state interpretation for the initial state indeed shows that Claims (1) - (3) hold:
	\begin{theorem}[Adequacy]
		Given an initial expression $\expr$, state $\state$, and cost $\cost$, some pure proposition $\varphi: Val \rightarrow Prop$, and potential postcondition $\potb$, the following holds for the execution of $\expr$:
		\begin{align*} 
		&\stateinterp(\state) * \wpP{\pota}{\expr}{v.~\potb(v), \varphi (v)} \rightarrow \\
		& \left(([\expr], \state, \cost) \tpstep^n \distr \rightarrow (v::\tpool', \state', \cost') \in \supp{\distr} \rightarrow \varphi(v)\right) \land &&\text{(postcondition)} \\
		& \left(([\expr], \state, \cost) \tpstep^n \distr \rightarrow (\tpool', \state', \cost') \in \supp{\distr} \rightarrow \expr' \in \tpool' \rightarrow \neg stuck(\expr',\state') \right) \land &&\text{(progress)} \\
		& \left(([\expr], \state, \cost) \tpstep^n \distr \rightarrow \pcost{\distr}{\potb} \leq \pota + \cost \right) &&\text{(expected cost)}
		\end{align*}
	\end{theorem}

\section{Case Studies} \label{sec:case-studies}

\subsection{Probabilistic Binary Counter}

In this section, we discuss a probabilistic variant of a classical example for amortized cost analysis: the binary counter.
In the deterministic variant of this example, one shows that the amortized number of bit flips needed for incrementing a binary counter (of any bit width) is in $\mathcal{O}(1)$, even though individual increments can be more expensive. We analyze a probabilistic version of the binary counter example, where a bit flip only succeeds with probability $p$. Each bit flip is repeated until it succeeds.

We analyze the following implementation of this example in ProbHeapLang. The counter is implemented by an array of size \lstinline{n}. The program iterates through the array from the least significant to the most significant bit, and flips every bit up to and including the first unset (0) bit.

%\begin{align*}
%  &\texttt{incrcounter n l p} \eqdef \\ 
%  & \qquad \Rec {\texttt{incr}} {\texttt{l i n}} =
%  & \qquad \If \texttt{i} = \texttt{n} then \TT 
%  & \qquad \Else \texttt{flipbit} (1 + \texttt{i}) \texttt{p};;
%\end{align*}

\begin{lstlisting}[language=mylang]
    incr_counter n l p := 
    (rec incr l i n :=
        if i = n then ()
        else flip_bit (l + i) p ;;
           if !(l + i)                // (l + i)th bit is now 1
           then ()
           else incr l (i + 1) n)
     l 0 n.
\end{lstlisting}

The implementation of \texttt{flip\_bit} is similar to the implementation of the coin toss example in Section \ref{sec:first-example}

\begin{lstlisting}[language=mylang]
    flip_bit :=
    rec flip l p :=
        Tick 1;;
        if ChooseWeighted [(p, true), (1-p, false)]
        then l <- not (!l)
        else flip l p.
\end{lstlisting}

Similar to the coin toss example, we can show that the expected cost needed for one bit flip is bounded by $1/p$.

We would like to show that m increments do not incur a cost larger than $\frac{2m}{p}$.

\begin{lstlisting}[language=mylang] 
    rec incr_m l m :=
        if m = 0 then ()
        else incr_counter n l p;; incr_m l (m-1)
\end{lstlisting}

\begin{theorem} \label{thm:prob_counter}
    Let $n \in \mathbb{N}_{+}, m \in \mathbb{N}, p \in \mathbb{R}$ with $0 < p < 1$. Then
  \upshape
\begin{align*}
|- \wpP{\left(\frac{2m}{p}\right)}{\Let \ell = {\mcode{init\_counter}~ n} in \mcode{incr\_m}~\ell~ m}{0, \TRUE}
\end{align*}
\end{theorem}

For proving this theorem, the key observation is that in each increment, we flip at most one bit from 0 to 1. We get a potential of $2/p$ for each increment, of which we use half, i.e., $1/p$, to ``pay'' for the cost of flipping the 0 to a 1, and we use the remaining $1/p$, to ``pay in advance'' for flipping the bit back to 0 eventually.

More precisely, we maintain a potential of $1/p$ for every set (1) bit in the counter. Whenever we flip a 1 to a 0, we ``pay'' for that flip using the potential that was associated with this bit. When we flip a $0$ to a $1$, we instead use $1/p$ of the additional potential to pay for the flip itself and associate the remaining $1/p$ with the new 1 in the counter. This strategy is captured by the following lemma.

\begin{lemma} \label{lem:prob_counter_ind}
    Let $p \in \mathbb{R}$, $0 < p < 1$, $\loc \in Loc$, $vs : list~ val$, and $n = length(vs)$. Let $B$ be a function that counts the number of ones in a list. Furthermore, define $is\_counter$ as $is\_counter~ \loc~ vs := \loc \mapsto* vs \text{ where vs is a boolean list}$. Then \upshape
    \begin{align*}
    &\left\{\left(\frac{2}{p} + \frac{1}{p} \cdot B(vs)\right) , is\_counter~ \loc~ vs \right\} \\
    &\qquad \qquad \mcode{incr\_counter}~ n~ \loc~ p \\ 
    &\left\{\lambda \_, \frac{1}{p} \cdot B(incr~ vs),is\_counter~ \loc~ (incr~ vs) \right\}  
    \end{align*}
\end{lemma}

The proof of Theorem \ref{thm:prob_counter} then proceeds by induction over the number of increments $m$, applying Lemma \ref{lem:prob_counter_ind} in the induction step.

\subsection{Quicksort}

Next, we use \logicName{} to bound the expected number of comparisons in an imperative in-place implementation of the quicksort algorithm. For simplicity, we assume that the list contains no duplicates.

The function \texttt{partition} reorders an array by putting all elements not greater than some pivot $x$ in the beginning, and all other elements to the end of the array.
\newpage
\begin{lstlisting}[language=mylang]
  partition n l x :=
    (rec partition i j :=
    if j = n then 0
      else 
        Tick 1;;
        if !(l + j) <= x
        then 
          swap (l + i) (l + j);;
          1 + partition (i + 1) (j + 1)
        else 
          partition i (j + 1)
    ) 0 0
\end{lstlisting}
\texttt{qsort} uses a derived construct \texttt{ChooseRange a b}, that models choosing uniformly from the integers from $a$ (inclusive) to $b$ exclusive. It can be defined using $\ChooseUniform{}$ on a list ranging from $a$ to $b-1$.
\begin{lstlisting}[language=mylang]
  qsort := 
    rec qsort l n :=
      if n <= 1 then () 
      else let i := ChooseRange 0 n in 
      let pos := partition l n !(l + i) in 
      qsort l pos;;
      qsort (l + pos) (n - pos)
\end{lstlisting}

We show a concrete bound of $cost(n) := 2n \cdot (1 + log_{4/3}(n))$ for the number of comparisons. 

\begin{theorem}[Expected Number of Comparisons for Quicksort] \label{thm:quicksort} \upshape
Let $vs$ be an integer list that contains no duplicates. Let $n=length(vs)$. Then
\begin{align*}
    \{\$cost(n), l \mapsto*~vs\}~ \textlang{qsort}~ l~n ~\{0, \TRUE\}
\end{align*}
\end{theorem}

The proof proceeds by induction on $n$. The key idea for the induction step is to distinguish ``good'' pivots, which lead to two sublists that are both significantly smaller than the input list, and ``bad'' pivots, which lead to highly unbalanced lists. We then show that the probability of choosing a good pivot is sufficiently large. This analysis is inspired by typical textbook proofs.

A good pivot is one that splits the list in a way that the larger sublist has at most a size of $\frac{3}{4}$th the size of the original list. 

\begin{lemma}[Good Pivot Split] ~\\
    Let $n > 0, \frac{1}{4} \cdot n \leq k \leq \frac{3}{4} \cdot n$. Then $cost(k) + cost(n-k) \leq cost(n) - 3n$.
\end{lemma}

\begin{lemma}[Bad Pivot Split] ~\\
    Let $0 < k \leq n$. Then $cost(k) + cost(n-k) \leq cost(n)$.
\end{lemma}

We know that \texttt{partition} incurs a cost of $n$ for a list of length $n$, and that the number of good pivots is at least $\lfloor \frac{n}{2} \rfloor$. For $n \geq 2$, we have that $\lfloor \frac{n}{2} \rfloor \geq \frac{n}{3}$. Therefore
\begin{align*} 
    \frac{1}{n} \sum_{k=1}^n n + cost(k) + cost(n-k) &\leq 
    n + \frac{1}{n} \left(\left\lfloor \frac{n}{2} \right\rfloor \cdot ( cost(n) - 3n ) + \left\lceil \frac{n}{2} \right\rceil \cdot cost(n) \right) \\
    &\leq n + cost(n) + \left\lfloor \frac{n}{2} \right\rfloor \cdot (-3) \\
    &\leq cost(n) 
\end{align*}

This allows us to distribute the potential of $cost(n)$ such that if the pivot is the $k$-th smallest element, we continue with a potential of $n + cost(k) + cost(n - k)$. Then, we use a potential of $n$ to pay for the cost of \texttt{partition}, the potential of $cost(k)$ to pay for the first recursive call, and the potential of $cost(n-k)$ to pay for the second recursive call. This concludes the proof of Theorem \ref{thm:quicksort}

\section{Limitations and Ongoing Work} \label{sec:limitations}

\logicName{}' choice of making the potential a parameter of the weakest precondition limits how potential can be used. A particular limitation is that it is impossible to make the postcondition potential depend on the final state that the program terminates in while adhering to standard Iris convention. Postconditions related to the state at some location $l$ are typically encoded in Iris as a proposition $\exists v.~l \mapsto v * \pred(v)$. However, expressing the postcondition potential in the same way is not possible, since the potential is not part of the logic. To understand why this is a problem, consider the following example, which is again a variant of the coin toss example: 

\begin{lstlisting}[language=mylang]
	toss l := if ChooseUniform [true, false] then ()
			else l <- !l + 1
	let l := alloc 0 in toss l;;Tick(!l)
\end{lstlisting}

We would like to show that the expected cost of this example is at most 2. For that, we would first have to show that if we execute \texttt{toss l} with an initial potential of $2$, we can distribute the potential such that if we terminate in a state where $l \mapsto v$, we have a potential of $v$ left. However, it is unfortunately impossible to express this postcondition in \logicName{}, since the potential is not part of the logic.

In ongoing work, we are exploring how to embed a general notion of \textit{probabilistic resources} in the base logic of Iris. We hope that making the potential a (probabilistic) resource in the logic rather than a parameter of the weakest precondition will enable the verification of examples like the one stated above.

\section{Related Work} \label{sec:related-work}

In concurrent work, Aguirre et al. \cite{Eris} introduce Eris, an Iris-based framework for reasoning about (expected) error bounds of probabilistic programs. The key novelty of Eris is \textit{error credits}, which can be distributed during the program execution to ``pay'' for outcomes that do not satisfy the postcondition. (Expected) error credits behave like the potential in our logic: When there is a probabilistic choice in the program, they can be distributed any way among the branches as long as the expected value does not increase. In contrast to our work, Aguirre et al. choose a different way to implement this behavior: whereas in our work, the potential is a parameter of the weakest precondition, they introduce a \textit{guarded lifting modality} that is used in the weakest precondition to distribute the credits. As a consequence, they do not encounter the limitation described in Section \ref{sec:limitations}, but are also not able to support concurrent programs.

Polaris \cite{Polaris} presents a different Iris-based approach for probabilistic reasoning. There, proofs proceed by first relating the program to an abstract model (using Iris), and then reasoning within that model about properties like expected cost. This is a very general technique for probabilistic reasoning. However, the technique requires finding a suitable abstract model. For expected cost analysis, the definition of an individual cost model is required for every proof.

Non-probabilistic runtime analysis in Iris was first studied through the concept of \emph{time credits} \cite{TimeCredits}.  \logicName{} was inspired by this work in that our potential behaves like time credits, but potential can be divided along probabilistic branching, which is a concept that does not arise in the deterministic setting of \cite{TimeCredits}.

A different, more general approach to probabilistic reasoning in separation logic relies on a modality that represents probabilistic conditioning \cite{Lilac, Bluebell}, in combination with an interpretation of the separating conjunction as probabilistic independence \cite{PSL}. However, all this work lacks fully modular reasoning (in the sense that there is no rule similar to \ruleref{wp-bind}), and the languages are sequential and first-order. We implicitly use probabilistic conditioning after every program step -- this is encoded in the definition of the weakest precondition (see \ref{sec:weakestpre-def}). In ongoing work, we are examining whether it is possible to support a similar proof strategy for \logicName{} using a conditioning modality. 

There is also a rich line of work on \emph{automated} amortized expected cost analysis \cite{ExpAuto, ExpAuto2} based on automated amortized resource analysis \cite{AARA1, AARA2}. Like our work, this line of work is also based on potentials. However, the focus is on automatically deriving polynomial bounds on expected costs of programs that make binary choices. Since this approach is designed for automated analysis, there is no support for correctness reasoning and it is, therefore, not suitable when the expected cost of a program depends on its functional behavior.

Another line of work develops Hoare-like calculi for expected cost analysis \cite{weakest-pre-expection,wpExp,AmortizedExp}, including amortized cost analysis, based on the method of potentials~\cite{AmortizedExp}. This line of work follows a syntax-based approach for computing the expected runtime of sequential first-order programs. Their support for correctness proofs is limited compared to Iris.

\bibliographystyle{plainnat}
\bibliography{thesis}

\begin{thebibliography}{24}
\providecommand{\natexlab}[1]{#1}
\providecommand{\url}[1]{\texttt{#1}}
\expandafter\ifx\csname urlstyle\endcsname\relax
  \providecommand{\doi}[1]{doi: #1}\else
  \providecommand{\doi}{doi: \begingroup \urlstyle{rm}\Url}\fi

\bibitem[Aguirre et~al.(2024)Aguirre, Haselwarter, de~Medeiros, Li, Gregersen,
  Tassarotti, and Birkedal]{Eris}
Alejandro Aguirre, Philipp~G Haselwarter, Markus de~Medeiros, Kwing~Hei Li,
  Simon~Oddershede Gregersen, Joseph Tassarotti, and Lars Birkedal.
\newblock Error credits: Resourceful reasoning about error bounds for
  higher-order probabilistic programs.
\newblock \emph{arXiv preprint arXiv:2404.14223}, 2024.

\bibitem[Bao et~al.(2024)Bao, D'Osualdo, and Farzan]{Bluebell}
Jialu Bao, Emanuele D'Osualdo, and Azadeh Farzan.
\newblock Bluebell: An alliance of relational lifting and independence for
  probabilistic reasoning.
\newblock \emph{CoRR}, abs/2402.18708, 2024.
\newblock \doi{10.48550/ARXIV.2402.18708}.
\newblock URL \url{https://doi.org/10.48550/arXiv.2402.18708}.

\bibitem[Barthe et~al.(2020)Barthe, Hsu, and Liao]{PSL}
Gilles Barthe, Justin Hsu, and Kevin Liao.
\newblock A probabilistic separation logic.
\newblock \emph{Proc. {ACM} Program. Lang.}, 4\penalty0 ({POPL}):\penalty0
  55:1--55:30, 2020.
\newblock \doi{10.1145/3371123}.
\newblock URL \url{https://doi.org/10.1145/3371123}.

\bibitem[Batz et~al.(2023)Batz, Kaminski, Katoen, Matheja, and
  Verscht]{AmortizedExp}
Kevin Batz, Benjamin~Lucien Kaminski, Joost{-}Pieter Katoen, Christoph Matheja,
  and Lena Verscht.
\newblock A calculus for amortized expected runtimes.
\newblock \emph{Proc. {ACM} Program. Lang.}, 7\penalty0 ({POPL}):\penalty0
  1957--1986, 2023.
\newblock \doi{10.1145/3571260}.
\newblock URL \url{https://doi.org/10.1145/3571260}.

\bibitem[Birkedal et~al.(2021)Birkedal, Dinsdale{-}Young, Gu{\'{e}}neau, Jaber,
  Svendsen, and Tzevelekos]{IrisExample4}
Lars Birkedal, Thomas Dinsdale{-}Young, Arma{\"{e}}l Gu{\'{e}}neau, Guilhem
  Jaber, Kasper Svendsen, and Nikos Tzevelekos.
\newblock Theorems for free from separation logic specifications.
\newblock \emph{Proc. {ACM} Program. Lang.}, 5\penalty0 ({ICFP}):\penalty0
  1--29, 2021.
\newblock \doi{10.1145/3473586}.
\newblock URL \url{https://doi.org/10.1145/3473586}.

\bibitem[Giry(2006)]{Giry}
Michele Giry.
\newblock A categorical approach to probability theory.
\newblock In \emph{Categorical Aspects of Topology and Analysis: Proceedings of
  an International Conference Held at Carleton University, Ottawa, August
  11--15, 1981}, pages 68--85. Springer, 2006.

\bibitem[Hoffmann et~al.(2011)Hoffmann, Aehlig, and Hofmann]{AARA2}
Jan Hoffmann, Klaus Aehlig, and Martin Hofmann.
\newblock Multivariate amortized resource analysis.
\newblock In Thomas Ball and Mooly Sagiv, editors, \emph{Proceedings of the
  38th {ACM} {SIGPLAN-SIGACT} Symposium on Principles of Programming Languages,
  {POPL} 2011, Austin, TX, USA, January 26-28, 2011}, pages 357--370. {ACM},
  2011.
\newblock \doi{10.1145/1926385.1926427}.
\newblock URL \url{https://doi.org/10.1145/1926385.1926427}.

\bibitem[Hofmann and Jost(2003)]{AARA1}
Martin Hofmann and Steffen Jost.
\newblock Static prediction of heap space usage for first-order functional
  programs.
\newblock In Alex Aiken and Greg Morrisett, editors, \emph{Conference Record of
  {POPL} 2003: The 30th {SIGPLAN-SIGACT} Symposium on Principles of Programming
  Languages, New Orleans, Louisisana, USA, January 15-17, 2003}, pages
  185--197. {ACM}, 2003.
\newblock \doi{10.1145/604131.604148}.
\newblock URL \url{https://doi.org/10.1145/604131.604148}.

\bibitem[Jung et~al.(2018)Jung, Krebbers, Jourdan, Bizjak, Birkedal, and
  Dreyer]{Iris}
Ralf Jung, Robbert Krebbers, Jacques{-}Henri Jourdan, Ales Bizjak, Lars
  Birkedal, and Derek Dreyer.
\newblock Iris from the ground up: {A} modular foundation for higher-order
  concurrent separation logic.
\newblock \emph{J. Funct. Program.}, 28:\penalty0 e20, 2018.
\newblock \doi{10.1017/S0956796818000151}.
\newblock URL \url{https://doi.org/10.1017/S0956796818000151}.

\bibitem[Kaiser et~al.(2017)Kaiser, Dang, Dreyer, Lahav, and
  Vafeiadis]{IrisExample2}
Jan{-}Oliver Kaiser, Hoang{-}Hai Dang, Derek Dreyer, Ori Lahav, and Viktor
  Vafeiadis.
\newblock Strong logic for weak memory: Reasoning about release-acquire
  consistency in iris.
\newblock In Peter M{\"{u}}ller, editor, \emph{31st European Conference on
  Object-Oriented Programming, {ECOOP} 2017, June 19-23, 2017, Barcelona,
  Spain}, volume~74 of \emph{LIPIcs}, pages 17:1--17:29. Schloss Dagstuhl -
  Leibniz-Zentrum f{\"{u}}r Informatik, 2017.
\newblock \doi{10.4230/LIPICS.ECOOP.2017.17}.
\newblock URL \url{https://doi.org/10.4230/LIPIcs.ECOOP.2017.17}.

\bibitem[Kaminski et~al.(2018)Kaminski, Katoen, Matheja, and Olmedo]{wpExp}
Benjamin~Lucien Kaminski, Joost{-}Pieter Katoen, Christoph Matheja, and
  Federico Olmedo.
\newblock Weakest precondition reasoning for expected runtimes of randomized
  algorithms.
\newblock \emph{J. {ACM}}, 65\penalty0 (5):\penalty0 30:1--30:68, 2018.
\newblock \doi{10.1145/3208102}.
\newblock URL \url{https://doi.org/10.1145/3208102}.

\bibitem[Krebbers et~al.(2017)Krebbers, Timany, and Birkedal]{IPM}
Robbert Krebbers, Amin Timany, and Lars Birkedal.
\newblock Interactive proofs in higher-order concurrent separation logic.
\newblock In Giuseppe Castagna and Andrew~D. Gordon, editors, \emph{Proceedings
  of the 44th {ACM} {SIGPLAN} Symposium on Principles of Programming Languages,
  {POPL} 2017, Paris, France, January 18-20, 2017}, pages 205--217. {ACM},
  2017.
\newblock \doi{10.1145/3009837.3009855}.
\newblock URL \url{https://doi.org/10.1145/3009837.3009855}.

\bibitem[Li et~al.(2023)Li, Ahmed, and Holtzen]{Lilac}
John~M. Li, Amal Ahmed, and Steven Holtzen.
\newblock Lilac: {A} modal separation logic for conditional probability.
\newblock \emph{Proc. {ACM} Program. Lang.}, 7\penalty0 ({PLDI}):\penalty0
  148--171, 2023.
\newblock \doi{10.1145/3591226}.
\newblock URL \url{https://doi.org/10.1145/3591226}.

\bibitem[Matsushita et~al.(2022)Matsushita, Denis, Jourdan, and
  Dreyer]{IrisExample5}
Yusuke Matsushita, Xavier Denis, Jacques{-}Henri Jourdan, and Derek Dreyer.
\newblock Rust{H}orn{B}elt: a semantic foundation for functional verification
  of {R}ust programs with unsafe code.
\newblock In Ranjit Jhala and Isil Dillig, editors, \emph{{PLDI} '22: 43rd
  {ACM} {SIGPLAN} International Conference on Programming Language Design and
  Implementation, San Diego, CA, USA, June 13 - 17, 2022}, pages 841--856.
  {ACM}, 2022.
\newblock \doi{10.1145/3519939.3523704}.
\newblock URL \url{https://doi.org/10.1145/3519939.3523704}.

\bibitem[McIver and Morgan(2005)]{weakest-pre-expection}
Annabelle McIver and Carroll Morgan.
\newblock \emph{Abstraction, Refinement and Proof for Probabilistic Systems}.
\newblock Monographs in Computer Science. Springer, 2005.
\newblock ISBN 978-0-387-40115-7.
\newblock \doi{10.1007/B138392}.
\newblock URL \url{https://doi.org/10.1007/b138392}.

\bibitem[M{\'{e}}vel et~al.(2019)M{\'{e}}vel, Jourdan, and
  Pottier]{TimeCredits}
Glen M{\'{e}}vel, Jacques{-}Henri Jourdan, and Fran{\c{c}}ois Pottier.
\newblock Time {C}redits and {T}ime {R}eceipts in {I}ris.
\newblock In Lu{\'{\i}}s Caires, editor, \emph{Programming Languages and
  Systems - 28th European Symposium on Programming, {ESOP} 2019, Held as Part
  of the European Joint Conferences on Theory and Practice of Software, {ETAPS}
  2019, Prague, Czech Republic, April 6-11, 2019, Proceedings}, volume 11423 of
  \emph{Lecture Notes in Computer Science}, pages 3--29. Springer, 2019.
\newblock \doi{10.1007/978-3-030-17184-1\_1}.
\newblock URL \url{https://doi.org/10.1007/978-3-030-17184-1\_1}.

\bibitem[Mulder and Krebbers(2023)]{IrisExample6}
Ike Mulder and Robbert Krebbers.
\newblock Proof automation for linearizability in separation logic.
\newblock \emph{Proc. {ACM} Program. Lang.}, 7\penalty0 ({OOPSLA1}):\penalty0
  462--491, 2023.
\newblock \doi{10.1145/3586043}.
\newblock URL \url{https://doi.org/10.1145/3586043}.

\bibitem[Ngo et~al.(2018)Ngo, Carbonneaux, and Hoffmann]{ExpAuto}
Van~Chan Ngo, Quentin Carbonneaux, and Jan Hoffmann.
\newblock Bounded expectations: resource analysis for probabilistic programs.
\newblock In Jeffrey~S. Foster and Dan Grossman, editors, \emph{Proceedings of
  the 39th {ACM} {SIGPLAN} Conference on Programming Language Design and
  Implementation, {PLDI} 2018, Philadelphia, PA, USA, June 18-22, 2018}, pages
  496--512. {ACM}, 2018.
\newblock \doi{10.1145/3192366.3192394}.
\newblock URL \url{https://doi.org/10.1145/3192366.3192394}.

\bibitem[O'Hearn(2007)]{SeparationLogic1}
Peter~W. O'Hearn.
\newblock Resources, concurrency, and local reasoning.
\newblock \emph{Theor. Comput. Sci.}, 375\penalty0 (1-3):\penalty0 271--307,
  2007.
\newblock \doi{10.1016/J.TCS.2006.12.035}.
\newblock URL \url{https://doi.org/10.1016/j.tcs.2006.12.035}.

\bibitem[Reynolds(2002)]{SeparationLogic2}
John~C. Reynolds.
\newblock Separation logic: {A} logic for shared mutable data structures.
\newblock In \emph{17th {IEEE} Symposium on Logic in Computer Science {(LICS}
  2002), 22-25 July 2002, Copenhagen, Denmark, Proceedings}, pages 55--74.
  {IEEE} Computer Society, 2002.
\newblock \doi{10.1109/LICS.2002.1029817}.
\newblock URL \url{https://doi.org/10.1109/LICS.2002.1029817}.

\bibitem[Sammler et~al.(2021)Sammler, Lepigre, Krebbers, Memarian, Dreyer, and
  Garg]{IrisExample3}
Michael Sammler, Rodolphe Lepigre, Robbert Krebbers, Kayvan Memarian, Derek
  Dreyer, and Deepak Garg.
\newblock Refined{C}: automating the foundational verification of {C} code with
  refined ownership types.
\newblock In Stephen~N. Freund and Eran Yahav, editors, \emph{{PLDI} '21: 42nd
  {ACM} {SIGPLAN} International Conference on Programming Language Design and
  Implementation, Virtual Event, Canada, June 20-25, 2021}, pages 158--174.
  {ACM}, 2021.
\newblock \doi{10.1145/3453483.3454036}.
\newblock URL \url{https://doi.org/10.1145/3453483.3454036}.

\bibitem[Swasey et~al.(2017)Swasey, Garg, and Dreyer]{IrisExample1}
David Swasey, Deepak Garg, and Derek Dreyer.
\newblock Robust and compositional verification of object capability patterns.
\newblock \emph{Proc. {ACM} Program. Lang.}, 1\penalty0 ({OOPSLA}):\penalty0
  89:1--89:26, 2017.
\newblock \doi{10.1145/3133913}.
\newblock URL \url{https://doi.org/10.1145/3133913}.

\bibitem[Tassarotti and Harper(2019)]{Polaris}
Joseph Tassarotti and Robert Harper.
\newblock A separation logic for concurrent randomized programs.
\newblock \emph{Proc. {ACM} Program. Lang.}, 3\penalty0 ({POPL}):\penalty0
  64:1--64:30, 2019.
\newblock \doi{10.1145/3290377}.
\newblock URL \url{https://doi.org/10.1145/3290377}.

\bibitem[Wang et~al.(2020)Wang, Kahn, and Hoffmann]{ExpAuto2}
Di~Wang, David~M. Kahn, and Jan Hoffmann.
\newblock Raising expectations: automating expected cost analysis with types.
\newblock \emph{Proc. {ACM} Program. Lang.}, 4\penalty0 ({ICFP}):\penalty0
  110:1--110:31, 2020.
\newblock \doi{10.1145/3408992}.
\newblock URL \url{https://doi.org/10.1145/3408992}.

\end{thebibliography}

\appendix

\addtocontents{toc}{\vspace{5mm}}

\chapter{Appendix}

\section{Full Weakest Precondition Definition} \label{wp-full}

    Changes to the Iris weakest precondition are highlighted in yellow.
	\begin{align*}
		\textdom{wp}(\stateinterp, \pred_F, \stuckness) \eqdef{}& \MU \textdom{wp\any rec}. \Lam \mask, \expr, \pred, \highlight{\pota, \potb}. \\
        & (\Exists\val. \toval(\expr) = \val \land \pvs[\mask] \pred(\val) \wedge \highlight{\pota \geq \potb(v))} \lor {}\\ 
		& \Bigl(\toval(\expr) = \bot \land \All \state, n_s, n_t. \stateinterp(\state, n_s, n_t)  \vsW[\mask][\emptyset] {} * \\
        &(s = \NotStuck \Ra \red(\expr, \state)) * \highlight{p \geq 0 * \All \distr. (\expr , \state , \step \distr ) \wand} \\
        &(\pvs[\emptyset]\later\pvs[\emptyset])^{n_\rhd(n_s)+1} \highlight{\exists \vec{\pota}.~\mathbb{E}_{(\_, \_, c) \sim \distr}[c] + \mathbb{E}_\distr[\vec{\pota}] \leq \pota} * \\
		&\highlight{\bigwedge\limits_{p'; (\expr', \state', \vec{\expr}, \_) \in \vec{p}; \supp \distr}
			\exists \pota', \vec{pt}.~ \pota'' + \sum \vec{pt} \leq \pota' ~ \wedge } \\
		&\qquad \pvs[\emptyset][\mask]  \stateinterp( n_s + 1, \state', n + |\vec{\expr}|) * \textdom{wp\any rec}(\mask, \expr', \pred, \highlight{\pota'', \potb}) * {}\\
		&\qquad \Sep_{\expr''; \pota''' \in \vec\expr; \vec{pt}} \textdom{wp\any rec}(\top, \expr'', \pred_F, \highlight{\pota''', 0})\Bigr) \\
		\wpP{\pota}[\stateinterp;\pred_F]{\expr}[\stuckness;\top]{\potb,\pred} \eqdef& ~\textdom{wp}(\stateinterp, \pred_F, \stuckness)~\expr~\pred~\pota~\potb \\
		\wpre[\stateinterp;\pred_F]{\expr}[\stuckness;\top]{\pred} \eqdef&
		 ~\wpP {0}[\stateinterp;\pred_F]{\expr}[\stuckness;top]{\mathbf{0},\pred}
	\end{align*}

\section{Full Adequacy Theorem} \label{adequacy_full}

\begin{theorem}[Adequacy]
    Assume we are given some $\vec\expr$, $\state$, $\cost$, $n$, $\distr$ such that $(\vec\expr, \state, \cost) \tpstep^n \distr$.
    Moreover, assume we are given , a stuckness parameter $\stuckness$ and \textit{meta-level} property $\metaprop$ that we want to show.
    To verify that $\metaprop$ holds, it is sufficient to show that there exist lists $\vec{\pota}, \vec{\potb}$ and $(\tpool_2, \state_2, \cost_2) \in \supp \distr$ such that the following Iris entailment holds:
    \begin{align*}
        &\proves \pvs[\top] \Exists \stateinterp, \vec\pred, \pred_F. \stateinterp(\state_1,0,0) * \left(\Sep_{\expr;\pred;\pota;\potb \in \vec\expr,\vec\pred;\vec{\pot},\vec{\potb}} \wpP{\pota}[\stateinterp;\pred_F]{\expr}[\stuckness;\top]{\potb,\pred}\right) * \\
        &\left(\consstate^{\stateinterp;\vec\pred;\pred_F}_{\stuckness;\vec{\pota};\vec{\potb};c}(\distr, \tpool_2, \state_2) \vs[\top][\emptyset] \hat{\metaprop}\right)
    \end{align*}
    where 
    \begin{align*}
        \consstate^{\stateinterp;\vec\pred;\pred_F}_{\stuckness;\vec{\pota};\vec{\potb};c}(\distr, \tpool_2, \state_2) \eqdef{}&\Exists \vec\expr_2, \tpool_2'. \tpool_2 = \vec\expr_2 \dplus \tpool_2' * {}\\
        &\quad |\vec\expr_1| = |\vec\expr_2| *{}\\
        &\quad (s = \NotStuck \Ra \All \expr \in \tpool_2. \toval(\expr) \neq \bot \lor \red(\expr, \state_2) ) *{}\\
        &\quad\exists n', n' \leq n.~ \stateinterp(\state_2, n', |\tpool_2'|) *{}\\
        &\quad \pcost{\distr}{\potb} \leq \cost + \sum \vec{\pota} ~ *\\
        &\quad \left(\Sep_{\expr,\pred \in \vec\expr_2,\vec\pred} \toval(\expr) \ne \bot \wand \pred(\toval(\expr))\right) *{}\\
        &\quad \left(\Sep_{\expr \in \tpool_2'} \toval(\expr) \ne \bot \wand \pred_F(\toval(\expr))\right)
    \end{align*}
    Where $\hat\metaprop$ embeds $\metaprop$ into \logicName.
\end{theorem}

\end{document}